# Multiplexed Multi-Color Raman Imaging of Live Cells with Isotopically Modified Single Walled Carbon Nanotubes

Zhuang Liu[†], Xiaolin Li[†], Scott M. Tabakman[†], Kaili Jiang[‡], Shoushan Fan[‡] and Hongjie Dai[†]*.

[†]*Department of Chemistry, Stanford University, Stanford, CA, 94305, USA.*

[‡]*Department of Physics and Tsinghua-Foxconn Nanotechnology Research Center, Tsinghua University, China*

We show that single walled carbon nanotubes (SWNTs) with different isotope compositions exhibit distinct Raman G-band peaks and can be used for multiplexed multi-color Raman imaging of biological systems. Cancer cells with specific receptors are selectively labeled with 3 differently 'colored' SWNTs conjugated with various targeting ligands including Herceptin (anti-Her2), Erbitux (anti-Her1) and RGD peptide, allowing for multi-color Raman imaging of cells in a multiplexed manner. SWNT Raman signals are highly robust against photo-bleaching, allowing long term imaging and tracking. With narrow peak features, SWNT Raman signals are easily differentiated from the auto-fluorescence background. The SWNT Raman excitation and scattering photons are in the near-infrared region, which is the most transparent optical window for biological systems in vitro and in vivo. Thus, SWNTs are novel Raman tags promising for multiplexed biological detection and imaging.

Fluorescence techniques have been widely used in biological imaging, though non-ideal factors exist including photo-bleaching of organic dyes, auto-fluorescence background from biological tissues, and wide fluorescence excitation and emission peaks giving rise to spectra overlays that limit the use of multiple colors in an experiment.[1] Raman scattering has narrow spectral lines that could be used for imaging with high multiplicity. In addition, tissue auto-fluorescence problems could be circumvented since sharp Raman peaks could be differentiated from fluorescence background. Raman excitation can also be chosen in low-background and biologically transparent optical windows. Raman imaging is promising for the next generation of biological imaging.[2-4]

Single walled carbon nanotubes (SWNTs)[2,5-13] and other nanomaterials[14,15] have been explored in biological and medical areas including drug delivery and imaging. As one dimensional quantum wires with sharp electronic density of states at the van-Hove singularities, SWNTs exhibit intrinsic optical properties including photoluminance in the NIR range[5,10] and strong resonant Raman scattering,[6,7] both of which have been used for biological imaging. Single-color SWNT Raman imaging of biological specimens has been studied previously,[2,6,7] but multi-color Raman imaging with SWNTs remains unexplored. Here, we show that SWNTs with different isotope compositions display well-shifted Raman G-band peaks,[16,17] and can serve as different colors for Raman imaging. Cancer cells with specific receptors are selectively tagged with 3 different 'color' SWNTs, allowing for multi-color Raman imaging of cells in a multiplexed manner.

Pure C12-SWNTs (C12-SWNT) were obtained commercially, while pure C13 SWNTs (C13-SWNT) and C12/C13 mixed SWNTs (C12/C13-SWNT) were synthesized by chemical vapor deposition (CVD) as previously reported[18] using C13-methane and mixed C12/C13 methane respectively. Note that CVD is a scalable and economic way for the synthesis of nanotubes, even with C13-methane. Phospholipid-PEG-amine (DSEP-PEG5000-Amine) functionalized C12-SWNT, C13-SWNT and C12/C13-SWNTs were conjugated to Herceptin (anti-Her2), Erbitux (anti-Her1) and arginine-glycine-aspartic acid (RGD) peptide to recognize Her2/neu, Her1/EGFR and integrin $\alpha_v\beta_3$ cell-surface receptors, respectively (Fig. 1a), following established protocols.[5,9] We chose these ligand-protein systems due to their importance in cancer biology and medicine. The Her1/EGFR and Her2/neu receptors belong to the

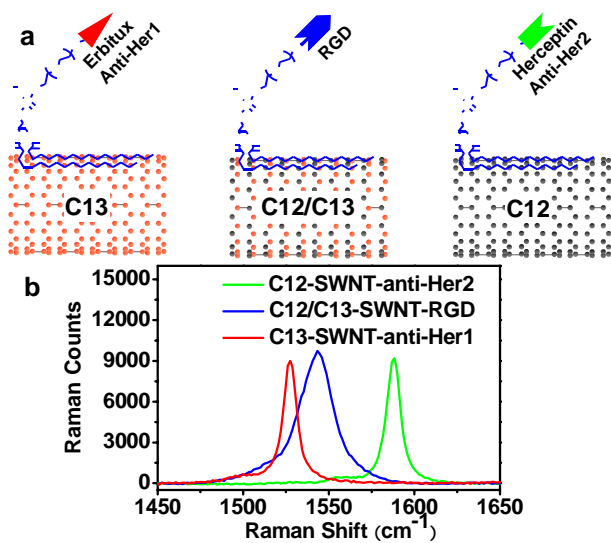

**Figure 1.** SWNTs with different Raman colors. (a) Schematic SWNTs with three different isotope compositions (C13-SWNT, C12/C13-SWNT, C12-SWNT) conjugated with different targeting ligands. (b) Solution phase Raman spectra of the three SWNT conjugates under 785 nm laser excitation. Different G-band peak positions were observed. At the same SWNT concentration, the peak height of C12-SWNT (Hipco) was ~2 times higher than that of C13-SWNT and ~4 times higher than that of C12/C13-SWNT. For mixtures used in biological experiments, concentrations of the three SWNTs were adjusted to give similar G-band peak intensities of the three colors, as shown in this figure.

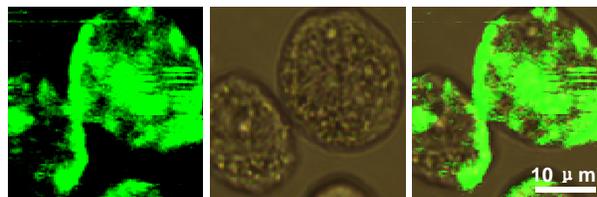

**Figure 2.** A confocal Raman spectroscopy image (left) of BT474 cells (optical image in middle, overlay image at right) after incubation with C12-SWNT-anti-Her2. Raman images were recorded with <1 μm spatial resolution (see SI).

ErbB protein family and are over-expressed on various cancer cells especially breast cancer cells.[19] Integrin $\alpha_v\beta_3$ is a cell surface receptor related to cancer angiogenesis and metastasis and is up-regulated on various solid tumor cells and fast growing tumor vasculatures.[20]

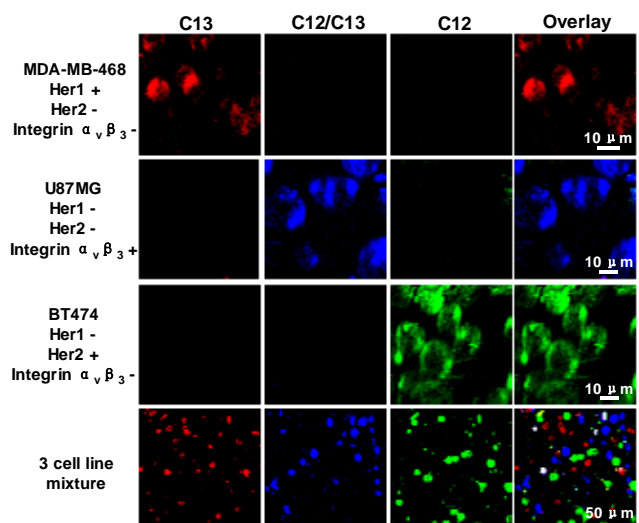

**Figure 3.** Multi-color Raman imaging with SWNTs. Deconvoluted confocal Raman spectroscopy images of three different cell lines after incubation with a mixture of the three-color SWNTs (top 3 rows, red, blue, green colors are Raman intensities of C12, C12/C13 and C13 SWNTs respectively). In the bottom row, a mixture of three cell lines was incubated with the three-color SWNT mixture. Those images clearly show a mixture of cells with differentiated Raman labeling by three types of SWNTs. Occasional co-localization of different colors could due to dead cells or non-specific nanotube binding.

Raman spectra of C12, C12/C13 and C13 SWNT conjugates were recorded under a near infrared (NIR) 785 nm laser excitation, displaying well-shifted Raman G-band peaks at 1590 cm$^{-1}$, 1544 cm$^{-1}$ and 1528 cm$^{-1}$ (Fig. 1b), as expected from the isotope effect.[16,17] We first incubated BT474 breast cancer cells (Her2 +) with C12-SWNT-anti-Her2 for 2 h at 4°C (to prevent endocytosis of nanotubes).[21] The cells were imaged by confocal Raman microscope after washing with phosphate buffered saline (PBS) for 3 times to remove non-specifically bound nanotubes on cells. Live cells were used directly for imaging. With 0.5 s collection time at each pixel, a full Raman spectroscopy image took 1~2 hours. This time could be shortened to ~20 min by using a higher power laser and adjusting the binning factor in the spectrometer. Association of SWNTs to the BT cells were clearly observed in high resolution confocal Raman spectroscopy images (Fig. 2). The targeting specificity was evidenced by that little SWNT signal was observed when cells were incubated with nanotubes without antibody conjugation (data not shown).

Next, we mixed the 3-color ligand-SWNTs for multiplexed cell imaging. Concentrations of the three SWNTs in the mixture were adjusted to give similar G-band peak intensities of the three colors. The peak shift from C12-SWNT to C13-SWNT was large without overlapping and can be easily separated (SI Fig. S1). The Raman spectrum of our three-color SWNT mixture (C12-SWNT-anti-Her2, C13-SWNT-anti-Her1, C12/C13-SWNT-RGD) however, showed overlaps between adjacent peaks. Deconvolution of a recorded Raman spectrum into the spectra of the three SWNT samples was done to analyze the relative amounts of each color in any mixture and on live cells.

Three types of cells including BT474 (human breast cancer, Her2+), U87MG (human glioblastoma, Integrin $\alpha_v\beta_3$+) and MDA-MB-468 (human breast cancer, Her1+) were incubated with the three-color SWNT mixture and then subjected to confocal Raman spectroscopy imaging. After background subtraction, the Raman spectrum recorded at each location ('pixel') was deconvoluted to intensities of the three colors (C12, C12/C13, C13) at each pixel, which was used to create images of the three colors representing the Raman signals of the different SWNTs (see SI for details). The resulting Raman spectroscopy images of three cell lines showed specific labeling of cells by SWNTs with minimal non-specific binding (Fig. 3, top 3 rows, SI Fig. S1). Furthermore, a mixture of the three cell lines was incubated with the three-color SWNT mixture. The Raman images clearly identified the existence of three cell types, each labeled by a distinct Raman color (Fig. 3, bottom row). This demonstrated the ability of multiplexed cell identification /imaging by SWNTs with different isotope compositions, for probing and imaging of several biological species simultaneously.

More SWNT Raman colors have been obtained previously by varying C12/C13 ratios in SWNTs.[17] Since the peak difference between C13-SWNT and C12-SWNT is 62 cm$^{-1}$ and that between C13-SWNT and C12/C13-SWNT is 16 cm$^{-1}$, more colors can be used for further multiplexing. The full width of half maximum (FWHM) of the SWNT G peak is < 2 nm, allowing for the SWNT Raman spectral features easily distinguished from auto-fluorescence background of cells. With a laser excitation of 785nm, the excitation and scattered photons are well within the NIR range (700-900 nm), which is in the most transparent optical window for biological imaging. Compared with other surface enhanced Raman scattering (SERS) nanoparticles,[2,4] SWNT Raman tags are simpler systems with well established synthesis methods.[18,22] The single isolated Raman G peak of SWNTs is also desirable. SWNTs exhibit high chemical inertness and stability, with robust Raman signals against photo-bleaching, which compares favorably over organic molecules.[7] This allows for tracking and imaging over long periods of time.[12] Beside the isotope-dependent Raman colors, SWNTs with controlled diameters also exhibit distinctive Raman peaks in their radial-breathing modes,[23] which can also be used for multi-color Raman imaging. Lastly, the various SWNT Raman colors can easily be excited with a single light source. Taken together, SWNTs are promising new Raman tags for multiplexed detection and imaging for biological systems.

**Acknowledgement:** We thank Drs. Sanjiv Gamhhir, Dean Felsher and Xiaoyuan Chen for providing us Herceptin, Erbitux and RGD, respectively. This work was supported by NIH-NCI CCNE-TR, NIH-NCI R01 CA135109-01 and a Stanford graduate fellowship.

**Supporting Information Available:** Experimental details and other supplementary data are available at http://pubs.acs.org.